\documentclass[3p,twocolumn]{elsarticle}

\usepackage{graphicx}
\usepackage{hyperref}       
\usepackage{url}                 
\usepackage{nicefrac}      
\usepackage{amsfonts, amsmath, amssymb}

\usepackage{color}
\usepackage{float}

\newcommand{\nn}{\nonumber \\}

\begin{document}

\title{Hadronic $J/\psi$ Regeneration in Pb+Pb Collisions}

\author{Joseph Dominicus Lap$^{1}$ and Berndt M\"uller$^{1,2}$}
\address{$^1$Department of Physics, Yale University, New Haven, CT 06511, USA}
\address{$^2$Department of Physics, Duke University, Durham, NC 27708, USA}

  
\begin{abstract}
We make use of published yields for $D$-mesons and $J/\psi$ in Pb+Pb collisions at ALICE and a schematic description of the expansion of the hadron gas to study $D$-meson collisions during the hadronic break-up phase as a production mechanism for charmonium in relativistic heavy ion collisions at the Large Hadron Collider. Our calculation is based on chemical reaction rates with thermal cross sections for an effective meson interaction among pseudoscalar and vector mesons. We find that due to regeneration, the newly measured $J/\psi$ yields are consistent with anywhere from roughly $25\%$ to $110\%$ of the total yield present at hadronization time. This allows us to bound the fractional abundance of $J/\psi$ immediately after hadronization: $0.28 \leq \frac{dN_0^{J/\psi}/dy}{dN_{\rm eq}^{J/\psi}/dy} \leq 1.13$. Our results are robust under the relaxation of the particulars of our schematic description and imply that it will be difficult to distinguish regeneration during hadronization from regeneration by final-state hadronic interactions. Therefore, regeneration must be taken into account when modelling.
\end{abstract}



\maketitle


\section{Introduction}

The surprisingly large yield of $J/\psi$ measured in Pb+Pb collisions at the Large Hadron Collider (LHC) \cite{ALICE:2012jsl,ALICE:2013osk} has been interpreted as evidence for the regeneration of $J/\psi$ mesons during hadronization of the quark-gluon plasma, which is oversaturated with charm \cite{Braun-Munzinger:2000csl,Thews:2000rj,Andronic:2003zv}. In the energy regime of the LHC ($\sqrt{s_{\rm NN}}=2.76, 5.02$ TeV) an abundant number of charm quark pairs is created during the initial collision of two $^{208}$Pb nuclei. Assuming sufficiently rapid thermalization, these $c$ and $\bar c$ quarks become an integral part of the quark-gluon plasma (QGP) that is created in the collision and eventually participate in the bulk hadronization of the QGP. 

Under the thermal conditions at hadronization ($T_H \sim 155$ MeV), the initially produced number of $c$ and $\bar c$ quarks is larger than the expected thermal yield and can be characterized by a fugacity factor $\lambda_c \approx 30$ \cite{Andronic:2021erx}. In the statistical hadronization model for charm (SHMc) this overpopulation of charm quarks results in a strong enhancement in the number of emitted charm quark containing hadrons, with $D$-mesons being enhanced by a factor $\lambda_c$ and charmonium states by a factor $\lambda_c^2$.

While the $D$-meson and $J/\psi$ yields predicted by the SHMc agree within uncertainties with the measured yields \cite{Andronic:2021erx,Andronic:2017pug,ALICE:2023gco}, $J/\psi$ production by coalescence of $c$ and $\bar c$ quarks at the QGP phase boundary may not the sole mechanism contributing to secondary $J/\psi$ production. For example, $J/\psi$ may be formed by recombination of $c$ and $\bar c$ quarks inside the QGP \cite{Thews:2000rj} once the temperature falls before the $J/\psi$ melting point (approximately $1.5T_c$ \cite{Datta:2003ww}). Charmonium can also be created by reactions among open charm hadrons in the late hadronic gas phase \cite{KO1998237,Zhang:2002ug,Bratkovskaya:2003ux,Linnyk:2008hp}. 

This means that the measured $J/\psi$ yield in the final state will be the result of at least four different production mechanisms: Primordial formation by gluon-gluon fusion, recombination of $c\bar{c}$ quarks within the QGP, recombination at the QGP-hadron phase boundary, and regeneration from $D/\bar{D}$-mesons in the hadron phase. In our study here, we explore to what extent the last of these sources can cloak the combined contributions from the first three. Is it possible from the data alone to conclude what the identifiable contribution from mechanisms other than regeneration in the hadronic phase to the measured yield is? If regeneration from $D/\bar{D}$ alone can establish chemical equilibrium, it is principally impossible to deduce the presence of earlier sources from the data. Since all previous studies of hadronic regeneration of $J/\psi$ \cite{KO1998237,Zhang:2002ug,Bratkovskaya:2003ux,Linnyk:2008hp} predate the availability of LHC data for $D$-meson production \cite{2022136986,ALICE:2021rxa} and $J/\Psi$ production\cite{ALICE:2023gco}; we here aim for an updated calculation that takes these data into account.

It is sometimes argued that after hadronization the chemical reaction rates among hadrons with open charm are too small to affect the number of produced $J/\psi$. However, this is not necessarily true as the overpopulation of charm quarks in the QGP also means that the produced number of mesons with open charm is much larger than what is expected in thermal equilibrium. This enhances the rates of $J/\psi$-producing reactions of the type $D+\bar{D} \to J/\psi+h$, where $h$ is a light hadron, quadratically. Furthermore, for many hadron states $h$, the regeneration reactions are exothermic and thus not suppressed as the hadron gas cools.

The total level of hadronic $J/\psi$ regeneration depends on the number of $D$-mesons present in the hadron gas – which can be measured – and on the magnitude of the cross sections for the hadronic charm exchange reactions, which are not known from experiment. This lack of knowledge naturally introduces a significant uncertainty of the calculated regeneration probability, which we will explore in Appendix A.

We will first briefly review the broken SU(4) symmetric model for $D$-meson–$J/\psi$ interactions and the reaction rates derived by Abreu {\em et al.} \cite{Abreu:2017cof}. We next discuss the rate equation governing $J/\psi$ regeneration and its initial conditions and describe several schematic expansion models. We then present our results for the $J/\psi$ yield and discuss their implications.

\section{Chemical Reaction Model} 

\subsection{$D$-meson--$J/\psi$ Interactions}

There exist several models for interactions among hadrons containing $c$ and $\bar c$ quarks.\footnote{An alternative example is the valence quark exchange model \cite{PhysRevC.51.2723,PhysRevC.62.045201}, which we do not consider here.} Here we consider the class of highly-constrained models \cite{Abreu:2017cof,Matinyan:1998cb,Haglin:1999xs,Lin:1999ad,PhysRevC.62.034903,Carvalho:2005et} that are based on an effective interaction among pseudoscalar ($P$) and vector ($V_\mu$) mesons with broken SU(4) flavor symmetry. The mesons are organized into multiplets in the adjoint representation of SU(4) for pseudoscalars:
\begin{equation}
P = \begin{pmatrix}
P_{+} & \pi^{+} & K^{+} & \bar{D}^{0}\\
\pi^- & P_{-} & K^{0} & D^{-}\\
K^{-} & \bar{K}^{0} & P_{3} & D_s^{-}\\
D^{0} & D^{+} & D_s^{+} & -\frac{3}{\sqrt{12}}\eta_c
\end{pmatrix} ,
\end{equation}
with
\begin{eqnarray}
    P_{\pm} &=& \pm\frac{\pi^0}{\sqrt{2}}+\frac{\eta}{\sqrt{6}} +\frac{\eta_c}{\sqrt{12}} , \nn
    P_{3} &=& -\frac{2\,\eta}{\sqrt{6}} +\frac{\eta_c}{\sqrt{12}} ,
\end{eqnarray}
and vector mesons:
\begin{equation}
V_\mu = \begin{pmatrix}
V_{+} & \rho^{+} & K^{*+} & \bar{D}^{*0}\\
\rho^- & V_{-} & K^{0*} & D^{*-}\\
K^{*-} & \bar{K}^{*0} & V_{3} & D_s^{*-}\\
D^{*0} & D^{*+} & D_s^{*+} & -\frac{3}{\sqrt{12}}J/\psi
\end{pmatrix} 
\end{equation}
with
\begin{eqnarray}
    V_{\pm} &=& \pm\frac{\rho^0}{\sqrt{2}}+\frac{\omega}{\sqrt{6}}+\frac{J/\psi}{\sqrt{12}} , \nn
    V_{3} &=& -\frac{2\,\omega}{\sqrt{6}} +\frac{J/\psi}{\sqrt{12}} .
\end{eqnarray}
Lorentz invariance and SU(4) symmetry then dictate the form of allowed interactions and determine vertices for the interactions of type PPV, PVV, VVV, PPPV, PPVV, PVVV, and VVVV. The relevant interactions are:
\begin{eqnarray}
\mathcal{L}_{\rm PPV} &=& -ig_{\rm PPV}\langle V^\mu[P,\partial_\mu P]\rangle \nn
\mathcal{L}_{\rm PVV} &=& g_{\rm PVV} \epsilon^{\mu\nu\alpha\beta} \langle\partial_\mu V_\nu\partial_\alpha V_\beta P\rangle \nn
\mathcal{L}_{\rm VVV} &=& ig_{\rm VVV}\langle\partial_\mu V_\nu[V^\mu,V^\nu]\rangle \nn
\mathcal{L}_{PPPV} &=& -ig_{\rm PPPV} \epsilon^{\mu\nu\alpha\beta}\langle V_\mu(\partial_\nu P)(\partial_\alpha P)(\partial_\beta P)\rangle \nn
\mathcal{L}_{\rm PPVV} &=& g_{\rm PPVV}\langle P V^\mu[V_\mu,P]\rangle \nn
\mathcal{L}_{\rm PVVV} &=& ig_{\rm PVVV} \epsilon^{\mu\nu\alpha\beta}\langle V_\mu V_\nu V_\alpha \partial_\beta P\rangle \nn
\mathcal{L}_{\rm VVVV} &=& g_{\rm VVVV}\langle V^\mu V_\nu[V^\mu,V^\nu]\rangle
\label{eq:rates}
\end{eqnarray}
The coupling constants for these interactions can be deduced from experimental data, and where such data are lacking, from SU(4) symmetry \cite{Matinyan:1998cb,Haglin:1999xs,Carvalho:2005et}.

\subsection{Reaction Rates}

Using these interactions one can calculate the isospin-averaged cross-section for $J/\psi$ absorption processes:
\begin{equation}
\sigma_r(s)=\frac{1}{64\pi^2s}\frac{p_f}{p_i}\int d\Omega \langle |\mathcal{M}_r(s,\theta)|^2 \rangle_{S,I},
\end{equation}
where $\mathcal{M}_r$ denotes the invariant amplitude for reaction $r$, $\sqrt{s}$ is the center of mass energy, $p_i$ ($p_f$) are the relative momenta in the initial (final) state of the reaction, and the average is taken over isospin and spin degrees of freedom weighted by their respective degeneracies. The matrix elements are usually calculated at lowest order in the meson coupling. The cross sections for inverse reactions can be obtained using the detailed balance relation
\begin{equation}
\sigma_{(3+4\to1+2)}=\sigma_{(1+2\to3+4)}
\frac{(2S_1+1)(2S_2+1)}{(2S_3+1)(2S_4+1)}\frac{p_i^2}{p_f^2} .
\end{equation}
Abreu {\it et al.}\cite{Abreu:2017cof} obtained estimates for the averaged cross-sections of various $J/\psi$ production processes. They also calculated the thermal average of the production cross-sections:
\begin{equation}
\langle \sigma_{ab\to cd}v_{ab} \rangle 
= \frac{\int d^3(p_a)d^3(p_b)f_a(p_a)f_b(p_b)\sigma_{ab\to cd}v_{ab}}
{\int d^3(p_a)d^3(p_b)f_a(p_a)f_b(p_b)}
\end{equation}
where $v_{ab}$ is the relative velocity of the incoming particles, and $f_i(p_i)$ is the temperature dependent Bose-Einstein distribution for particles of species $i$.

As the reactions are exothermic, the resultant thermally averaged $J/\psi$ production cross sections vary little over the relevant temperature range (90 MeV $ \leq T \leq$ 155 MeV). The following values are taken from Figs. 4,5 of \cite{Abreu:2017cof} using error bars to constrain any deviations from a constant value:
\begin{eqnarray}
\langle\sigma v\rangle(D\bar{D}\to J\Psi+\pi) &=& .03\pm.01~{\rm mb} \nn
\langle\sigma v\rangle(D^*\bar{D}^*\to J\Psi+\pi) &=& 1~{\rm mb} \nn 
\langle\sigma v\rangle(D^*\bar{D}\to J\Psi+\pi) &=& 3.75\pm.25~{\rm mb} \nonumber
\end{eqnarray}
\begin{eqnarray}
\langle\sigma v\rangle(D\bar{D}\to J\Psi+\rho) &=& .065\pm.01~{\rm mb} \nn 
\langle\sigma v\rangle(D^*\bar{D}\to J\Psi+\rho) &=& .15\pm.05~{\rm mb} \nn 
\langle\sigma v\rangle(D^*\bar{D}^*\to J\Psi+\rho) &=& .95\pm.05~{\rm mb} \nonumber
\end{eqnarray}
\begin{eqnarray}
\langle\sigma v\rangle(D_s\bar{D}\to J/\psi+K) &=& .25\pm.05~{\rm mb} \nn 
\langle\sigma v\rangle(D_s^*\bar{D}\to J/\psi+K) &=& 1.7~{\rm mb} \nn 
\langle\sigma v\rangle(D_s\bar{D}^*\to J/\psi+K) &=& 1.5~{\rm mb} \nn 
\langle\sigma v\rangle(D_s^*\bar{D}^*\to J/\psi+K) &=& .4~{\rm mb} \nonumber
\end{eqnarray}
\begin{eqnarray}
\langle\sigma v\rangle(D_s\bar{D}\to J\Psi+K^*) &=& 0.3\pm0.1~{\rm mb} \nn
\langle\sigma v\rangle(D_s^*\bar{D}\to J\Psi+K^*) &=& .8~{\rm mb} \nn 
\langle\sigma v\rangle(D_s\bar{D}^*\to J\Psi+K^*) &=& 1.2~{\rm mb} \nn 
\langle\sigma v\rangle(D_s^*\bar{D}^*\to J\Psi+K^*) &=& 4~{\rm mb} \nonumber
\end{eqnarray}

We note that the thermal cross sections estimated by Bratkovskaya {\it et al.} \cite{Bratkovskaya:2003ux} in their phenomenological model agree quite well with those obtained by Abreu {\it et al.} \cite{Abreu:2017cof} in the effective SU(4) chiral model. Before moving on, it is worth mentioning that Abreu {\it et al.} also considered reactions involving the $s$-channel $Z_c(3900)$ and $Z_c(4025)$ resonances and found them to be small enough to safely ignore.\\
For the absorption cross-sections we use the full temperature dependent expressions. However, as we note in Appendix A the final yield is fairly insensitive to the temperature dependence of the cross-sections.

\section{Rate Equation}

The time evolution of the number of $J/\psi$ particles, $N_{J/\psi}$, is described by the following rate equation \cite{Abreu:2017cof}:
\begin{eqnarray} 
\frac{dN_{J/\psi}(\tau)}{d\tau} &=& \sum_{p_1,p_2,h} \langle\sigma_{p_1+p_2\to J/\psi+h}\rangle \frac{N_{p_1}(\tau)N_{p_2}(\tau)}{V(\tau)} 
\nn 
& & -\sum_{p_1,p_2,h} \langle\sigma_{J/\psi+h\to p_1+p_2}\rangle \frac{N_{J/\psi}(\tau)N_{h}(\tau)}{V(\tau)} , 
\label{eq:rate}
\end{eqnarray}
where $N_{p_1},N_{p_2}$ denote the abundances of the various species of $D$-mesons and $N_h$ denotes the number of light hadrons in the initial state of the reverse reaction. Here we assumed that the hadrons are evenly distributed over the density profile of the fireball. This assumption allows us to write the rate equations for volume integrated numbers of mesons $N_i$ instead of local particle densities $n_i$. The first term in (\ref{eq:rate}) accounts for formation reactions; the second term accounts for absorption processes. Assuming boost invariance – an assumption that is well justified at midrapidity in heavy ion collisions in the LHC energy range – we will use particle numbers and volume per unit rapidity, $dN_i/dy$ and $dV/dy$, but denote these simply as $N_i$ and $V$.

It is useful to rewrite (\ref{eq:rate}) in the following form:
\begin{equation}
\frac{dN_{J/\psi}(\tau)}{d\tau} = \frac{C_{J/\psi}- A_{J/\psi}(\tau) N_{J/\psi}(\tau)}{V(\tau)} 
\label{eq:prodrate}
\end{equation}
where
\begin{eqnarray}
C_{J/\psi} &=& \sum_{p_1,p_2,h}\langle\sigma_{p_1+p_2\to J/\psi+h}\rangle N_{p_1}N_{p_2} ,
\\
A_{J/\psi}(\tau) &=& \sum_{p_1,p_2,h} \langle\sigma_{J/\psi+h\to p_1+p_2}\rangle (\tau)N_{h}.
\end{eqnarray}
Because the thermal production cross sections are nearly temperature independent and the number of $D$-mesons does not appreciably change during the hadronic expansion stage, $C_{J/\psi}$ remains approximately constant during the entire hadronic fireball expansion. On the other hand, $A_{J/\psi}(\tau)$ drops rapidly with time as the hadronic gas expands and cools, because the thermal $J/\psi$ absorption cross sections decrease rapidly with falling temperature and the light hadron densities drop inversely with the volume of the fireball.\\

The solution for yields as a function of time is then:
\[N_{J/\psi}(\tau)=\frac{N_{J/\psi}(\tau_H)+C_{J/\psi}\int_{\tau_H}^\tau \frac{\mu(\tau')}{V(\tau')}d\tau'}{\mu(\tau)}\]
Where $\mu(\tau)=\exp \left( \int_{\tau_H}^\tau \frac{A_{J/\psi}(\tau')}{V(\tau')}d\tau' \right)$.
\subsection{Initial conditions}

We base our investigation on the measured yields $dN_i/dy$ for $D$-mesons in $0-10$\% central Pb+Pb collisions at $\sqrt{s_{\rm NN}}=5.02$ TeV (see Table 4 in \cite{ALICE:2021rxa} and Table 3 of \cite{2022136986}). Because the system is almost net-baryon free, particle and antiparticle yields are assumed to be identical. The $D^0$ yield listed in \cite{ALICE:2021rxa} includes $D^{0*}$, and the $D_s$ yield in \cite{2022136986} includes the $D_s^{*}$. In order to estimate the separate yields for these species we assume the yields are in the same proportion as the identified $D^{+}/D^{+*}$ ratio ($0.80+.69-.36$). We also assume that the fractional uncertainty on the subdivided results is maintained. This gives the following separate yield estimates:
\begin{eqnarray}
N(D^0) &=& 3.03^{+.63}_{-.64} \nn
N(D^{0*}) &=& 3.79^{+.79}_{-.80} \nn
N(D^+) &=& 3.04^{+.63}_{-.90} \nn
N(D^{*+}) &=& 3.80^{+1.12}_{-1.34} \nn
N(D_s) &=& 0.84^{+.24}_{-.38} \nn
N(D_s^{*}) &=& 1.05^{+.35}_{-.47}
\label{eq:ND}
\end{eqnarray}

These numbers include feed-down from multiple short-lived states with $D$-meson quantum numbers. Estimates indicate that up to 75\% of the measured yields (\ref{eq:ND}) originate from the decay of such states. Since most of the higher lying $D$-meson states in the Particle Data Book \cite{ParticleDataGroup:2020ssz} have decay widths greater than 200 MeV,\footnote{the $D_2^*(2460)$ with a width of 47 MeV is an exception} corresponding to lifetimes less than 1 fm/c, we will here assume that these feed-down processes are complete by the time when we begin integrating the rate equation (\ref{eq:prodrate}). If some of the higher excited states (such as the $D_2^*(2460)$) remain populated, they would also contribute to $J/\psi$ regeneration. While we do not know what the corresponding thermal cross sections are, it is a reasonable assumption that they would be comparable to the cross sections (\ref{eq:rates}) for $J/\psi$ production from the low-lying $D$-meson states. We will therefore ignore this complication and simply use the yields listed in (\ref{eq:ND}).

Using these values for the $D$-meson yields immediately after hadronization of the QGP and the thermally averaged cross sections the constant prefactor $C_{J/\psi}$ in (\ref{eq:prodrate}) takes the value
\begin{equation}
C_{J/\psi} = 53^{+28}_{-33} ~{\rm fm}^2.
\end{equation}
The uncertainty range accounts for the uncertainty in the number of $D$-mesons formed during hadronization, which enters quadratically into $C_{J/\psi}$.

\subsection{Expansion Models}

The final missing piece in our calculation is the time evolution of the volume $V(\tau)$ of the fireball per unit rapidity. As the thermal cross sections for $J/\psi$ production are not known from experiment and have large theoretical uncertainties, it makes sense to adopt a schematic model for the expansion of the hadronic system instead of conducting a detailed transport simulation of its expansion. As already stated, we assume that the hadrons are uniformly distributed over a large volume at the moment of hadronization and maintain this uniformity during the expansion with a prescribed velocity profile. 

Both Abreu {\it et al.} \cite{Abreu:2017cof} and Andronic {\it et al.} \cite{Andronic:2021erx} model the expanding hadronic system as a cylindrical fireball, but their schematic models differ in details. Both models assume a global hadronization proper time $\tau_H$, at which the QGP converts into hadrons. Here we compare the two models in order to get a sense of how sensitive the produced $J/\psi$ yield is to those details.

Abreu {\it et al.} \cite{Abreu:2017cof} estimate the fireball volume as a function of the global proper time $\tau$ by:
\begin{equation}
V(\tau) = c\tau \pi R(\tau)^2 ,
\end{equation}
where $R(\tau)$ denotes the expanding transverse radius of the fireball:
\begin{equation}
R(\tau) = R_H+v_H(\tau-\tau_H)+\frac{a_H}{2}(\tau-\tau_H)^2 .
\end{equation}
We adapt the parameter values given in \cite{Abreu:2017cof} for Au+Au collisions at RHIC to Pb+Pb collisions at LHC. We choose these values such that they correspond to a hadronization volume $V(\tau_H) = 4,997\pm455~{\rm fm}^3$ \cite{Andronic:2021erx}.

Andronic {\it et al.} \cite{Andronic:2021erx} calculate the hadronization volume at chemical freeze-out as a one-dimensional integral along the freeze-out contour in the $\tau-r$ plane, where $\tau$ is the local proper time and $r$ denotes the transverse coordinate:
\begin{equation}
V_H = 2\pi\int_0^{R_H} dr\, r\, \tau(r)\, u^\tau\, [1-\beta(r)\frac{\partial\tau}{\partial r}] .
\end{equation}
where $u^\tau=(1-\beta^2)^{-1/2}$ is the Lorentz factor associated with transverse expansion and $\beta(r)$ is the radial velocity. This approach enables them to study different freeze-out surfaces $\tau(r) = R_H+\int_0^rdr'\beta(r')$ with a transverse flow profile $\beta(r)$. We adopt the scaling model $\beta(r) = \beta_H(r/R_H)^n$ with values $n=0.85$ and $\beta_H=0.62$ as advocated in \cite{Andronic:2021erx} and $n=0.83$, $\beta_H=0.57$ as advocated in \cite{Mazeliauskas:2019ifr}. By requiring that both models have same volume at hadronization we can calculate a hadronization radius $R_H$ and time $\tau_H$. The uncertainty of the volume corresponds to the range $10.15~{\rm fm} \leq R_H \leq 12.25~{\rm fm}$, which includes the transverse radius $R_H=12~{\rm fm}$ used in the Abreu {\it et al.} model. Based on this hadronization radius we find a hadronization time of $9.63~{\rm fm}/c \leq\tau_H\leq 16.84~{\rm fm}/c$ for the Abreu model (instead of using the values  $v_H=0.6,a_H=.044$ estimated in Ref.~\cite{Abreu:2017cof} we use the values $v_H=0.62,a_H=0$ from Andronic {\it et al.} \cite{Andronic:2021erx}). We find much tighter bounds of $10.85~{\rm fm}/c \leq\tau_H\leq 11.53~{\rm fm}/c$ and $10.15~{\rm fm}/c \leq\tau_H\leq 10.79~{\rm fm}/c$ for the Andronic {\it et al.} \cite{Andronic:2021erx} and Mazeliauskas {\it et al.} \cite{Mazeliauskas:2019ifr} volume models respectively.

In order to estimate the freeze-out volume $V_F$, we use the relation 
\begin{equation}
\frac{V_F}{V_H} \geq \frac{\gamma_H}{\gamma_F}\left(\frac{T_H}{T_F}\right)^3 
\label{eq:VF}
\end{equation}
where $\gamma_H$ and $\gamma_F$ are the fugacities at hadronization and freeze-out and equality holds for isentropic expansion (which we'd expect from a gas of massless particles). As Xu and Ko \cite{Xu_2017} have shown, the entropy per particle stays roughly constant during the hadronic expansion, and thus we do not expect large deviations from equality. In addition the fugacities are for light quarks and we assume they are approximately unity. Our results are however insensitive to the assumption of isentropic expansion and unit fugacities. If one remains agnostic as to the temperature dependence of the cross sections (thereby remaining agnostic as to the form of the expansion and fugacities) one obtains consistent results albeit with larger error bars\footnote{It is worth pointing out that we will find that regeneration continues to occur until $\tau\rightarrow\infty$ rather than ending at the freeze-out time $\tau_F$. Therefore, the value of the freeze-out volume obtained by assuming equality in (\ref{eq:VF}) does not factor into our main result.}. The hadronization temperature and freeze-out temperature are estimated to be $T_H \approx 155$ MeV and $T_F \approx 90$ MeV, respectively \cite{ALICE:2013mez}.This yields $17~{\rm fm}/c\lessapprox\tau_F \lessapprox29~{\rm fm}/c$ and $V_F \approx 25,000~{\rm fm}^3$.

Normally one evolves the hadronic fireball until the thermal freeze-out time when the elastic scattering among hadrons becomes ineffective. However, as the $D$-meson reactions leading to $J/\psi$ production are exothermic and thus $\langle \sigma v \rangle$ remains constant even after nominal freeze-out, there is no justification for stopping the calculation at the freeze-out time. The production of $J/\psi$ merely becomes less common by dilution as the volume increases. As the integral $\int d\tau/V(\tau)$ is convergent, it makes sense to simply integrate until $\tau\to\infty$ rather than imposing an artificial upper time limit.

\begin{figure}
    \centering
    \includegraphics[width=0.8\linewidth]{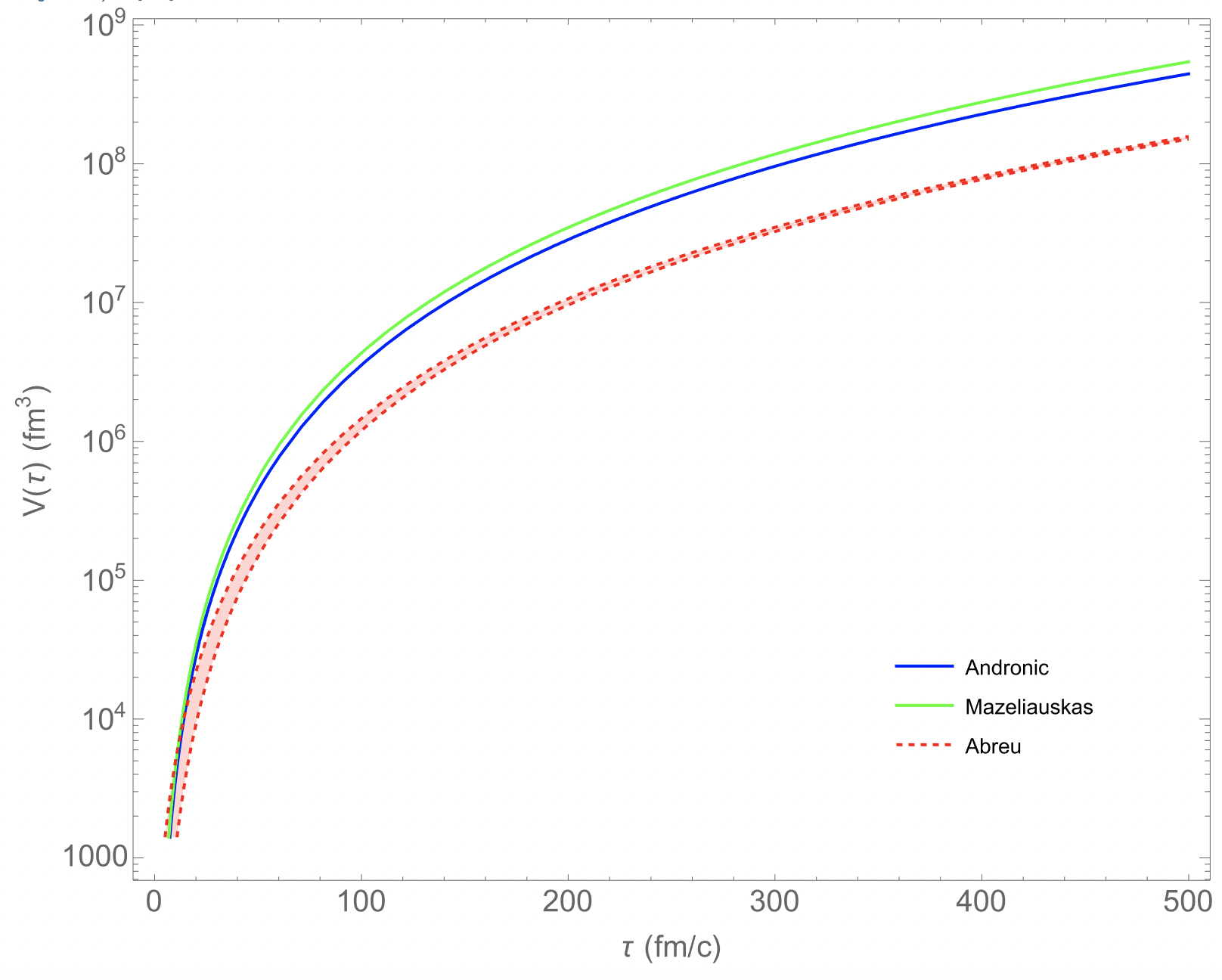}
    \caption{Comparison of volume evolution post hadronization for Bjorken freeze-out surface}
    \label{fig:VolComp}
\end{figure}

We find that the physics is not very sensitive to the details of the expansion models as shown in Fig.~\ref{fig:VolComp}. For example, we have used the different values for $n$ and $\beta$ given in \cite{Andronic:2021erx} and \cite{Mazeliauskas:2019ifr} but found that the behaviour is insensitive to both parameters within the error bars. While we have taken into account the temperature dependence of the cross sections this required us to enforce isentropic expansion. However, if one remains agnostic concerning the details of the expansion process considering instead just the upper and lower bound of the cross sections, rather than their temperature dependence, one finds the total yields are also insensitive to this choice, but the error bars are increased as one would expect.

\section{Results}

For a first exploration of the importance of hadronic regeneration of $J/\psi$, we make the extreme assumption that no $J/\psi$ is present immediately after hadronization, i.~e.\ $N_{J/\psi}(\tau_H)=0$. 

\begin{figure}[h]
    \centering
    \includegraphics[width=0.8\linewidth]{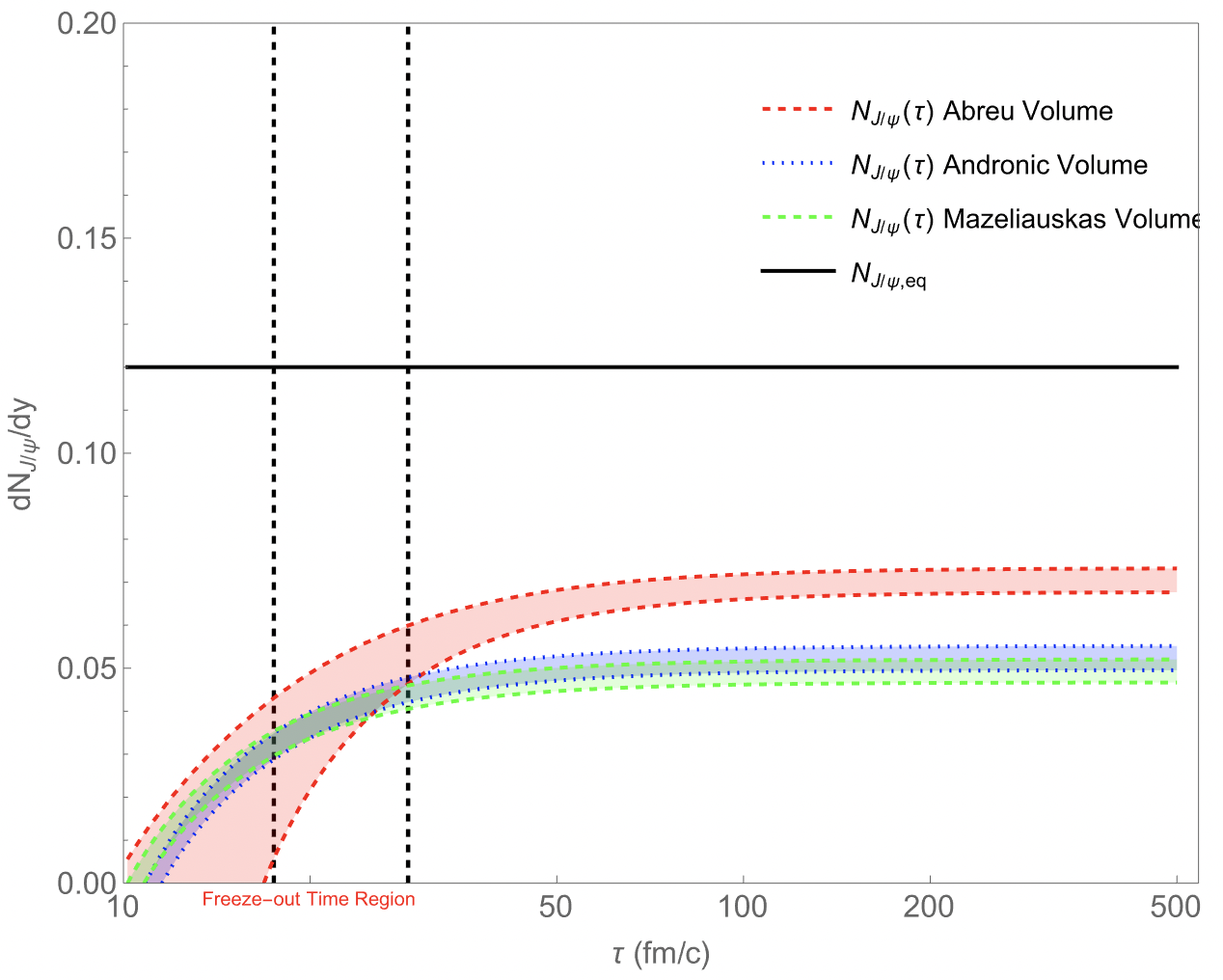}
    \caption{Time dependence of hadronic $J/\psi$ production by $D$-mesons. We note that $D$-meson interactions occurring after the estimated kinetic freeze-out time, $\sim 20$ fm/$c$, contribute significantly to the total $J/\psi$ yield. The horizontal black line indicates the chemical equilibrium yield. N.B. These yields are provided without uncertainty due to $D$-meson yields for legibility.}
    \label{fig:N_Jpsi}
\end{figure}

The time dependence of $N_{J/\psi}$ for the three different evolution models is shown in Fig.~\ref{fig:N_Jpsi}. The number of $J/\psi$ converges to an asymptotic value in both cases, but the yield still increases substantially after the nominal freeze-out time $\tau_F$, and it takes until $\tau\sim 50~{\rm fm}/c$ for the yield to remain constant with time. Even though $D$-meson collisions becoming increasingly rare as the hadron gas expands, terminating the calculation at the thermal freeze-out time would underestimate the total yield. 

\begin{figure}[H]
    \centering
    \includegraphics[width=0.8\linewidth]{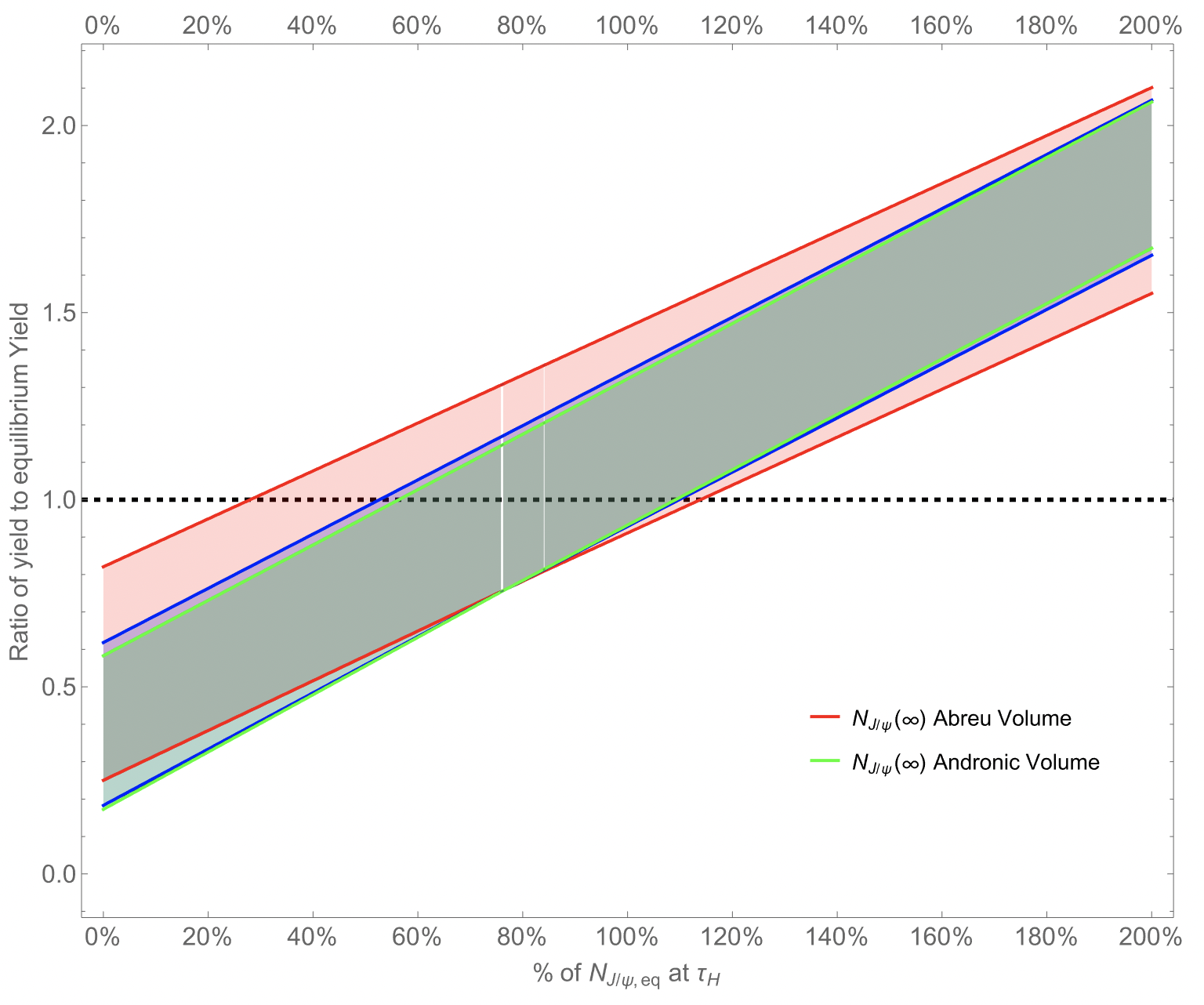}
    \caption{Ratio of total $J/\psi$ yield at $\tau=+\infty$ to equilibrium yield including the effect of absorption and assuming a fraction of the equilibrium yield of $J/\psi$ (shown on the abscissa) is produced during hadronization.}
    \label{fig:Percent}
\end{figure}

Figure~\ref{fig:N_Jpsi} shows that the final $J/\psi$ yield from $D$-meson reactions starting from zero at $\tau_H$, is $0.047<N_{J/\psi}<0.073$. This is a significant fraction of the measured equilibrium $J/\psi$ yield of $N_{\rm exp}(J/\psi) = 0.12\pm.017$ \cite{ALICE:2023gco} and it is reasonable to wonder how much hadronically produced $J/\psi$ is necessary to arrive at the equilibrium yield. As such, we now study how the total $J/\psi$ yield changes if we vary the number of $J/\psi$ assumed to be present immediately after hadronization. The results are shown in Fig.~\ref{fig:Percent} as a function of the equilibrium fraction of $J/\psi$ at $\tau_H$.  As expected, starting from a larger number of $J/\psi$ increases the final $J/\psi$ yield, but it is clear from Figure \ref{fig:Percent} that a wide range of initial conditions are compatible with the measured yields. Such an effect is irrespective of the schematically chosen volume models.

\section{Summary and Discussion}

Our calculation demonstrates that we cannot safely ignore the contribution from hadronic charmonium production processes to the $J/\psi$ yield in heavy-ion collisions at LHC energies. As such, any calculation of thermal production of $J/\psi$ must take regeneration by $D$-meson collisions into account. The central result of our study is that the measured $J/\psi$ yield permits us to set lower and upper bounds on the fractional abundance of $J/\psi$ immediately after hadronization:
\begin{equation}
0.28 \leq \frac{dN_0^{J/\psi}/dy}{dN_{\rm eq}^{J/\psi}/dy} \leq 1.13 \, .
\label{eq:result}
\end{equation}
These bounds could be further tightened by improving the expansion model or by reducing the uncertainty of the measured $D$-meson and $J/\psi$ yields.

We expect our results to underestimate the true yield from hadronic regeneration of $J/\psi$ as certain production channels have been neglected, such as $D_s+\bar{D}_s\to J/\psi+\phi$. A non-uniform distribution of $c$ ($\bar c$) quarks with greater concentration toward the center of the collision, as is expected from the transverse distribution of binary nucleon-nucleon collisions, also tends to increase the overall yield. Finally, it is unknown what fraction of $D\bar D$ collisions result in the formation of excited charmonium states, especially the $\chi_1$, which then would produce additional $J/\psi$ via feed-down. There is no reason to expect that the primary ratio $N(\chi_1)/N(J/\psi)$ in $D$-meson mediated reactions is given by the ratio obtained in the SHMc.

Since these additional sources of uncertainty tend to enhance the $J/\psi$ yield we surmise that our result is likely an underestimate. We therefore conclude that $D$-meson reactions contribute substantially to the measured $J/\psi$ yield in Pb+Pb collisions at the LHC and must be taken into account in microscopic transport models of the final hadronic stage of the heavy-ion reaction.  As such, it is not possible to conclude that the observed equilibrium yield of $J/\psi$ mesons provides unambiguous evidence for the formation of all observed $J/\psi$ during the hadronization transition.

{\it Acknowledgement:} We thank A.~Andronic, E.~Bratkovskaya, P.~Braun-Munzinger, and J.~Stachel for valuable comments on a draft of this manuscript. This work was supported by grant DE-FG02-05ER41367 from the Office of Science of the U.S. Department of Energy. B.M. acknowledges hospitality and financial support during a sabbatical stay at Yale University.
\bigskip

\section*{Appendix A: Robustness}

In our study of how regeneration obfuscates the degree of the $J/\psi$ yield we were forced to make some choices: assuming isentropic expansion, and using an effective model to calculate the production and absorption cross-sections. One might reasonably wonder how robust this effect is, that is whether it is an artifact of the assumptions we've made. The purpose of this section is to soothe the skeptical reader by showing that relaxing these assumptions reproduces the same result albeit with larger error bars.
\subsection{Isentropic expansion}
\begin{figure}[H]
    \centering
    \includegraphics[width=0.8\linewidth]{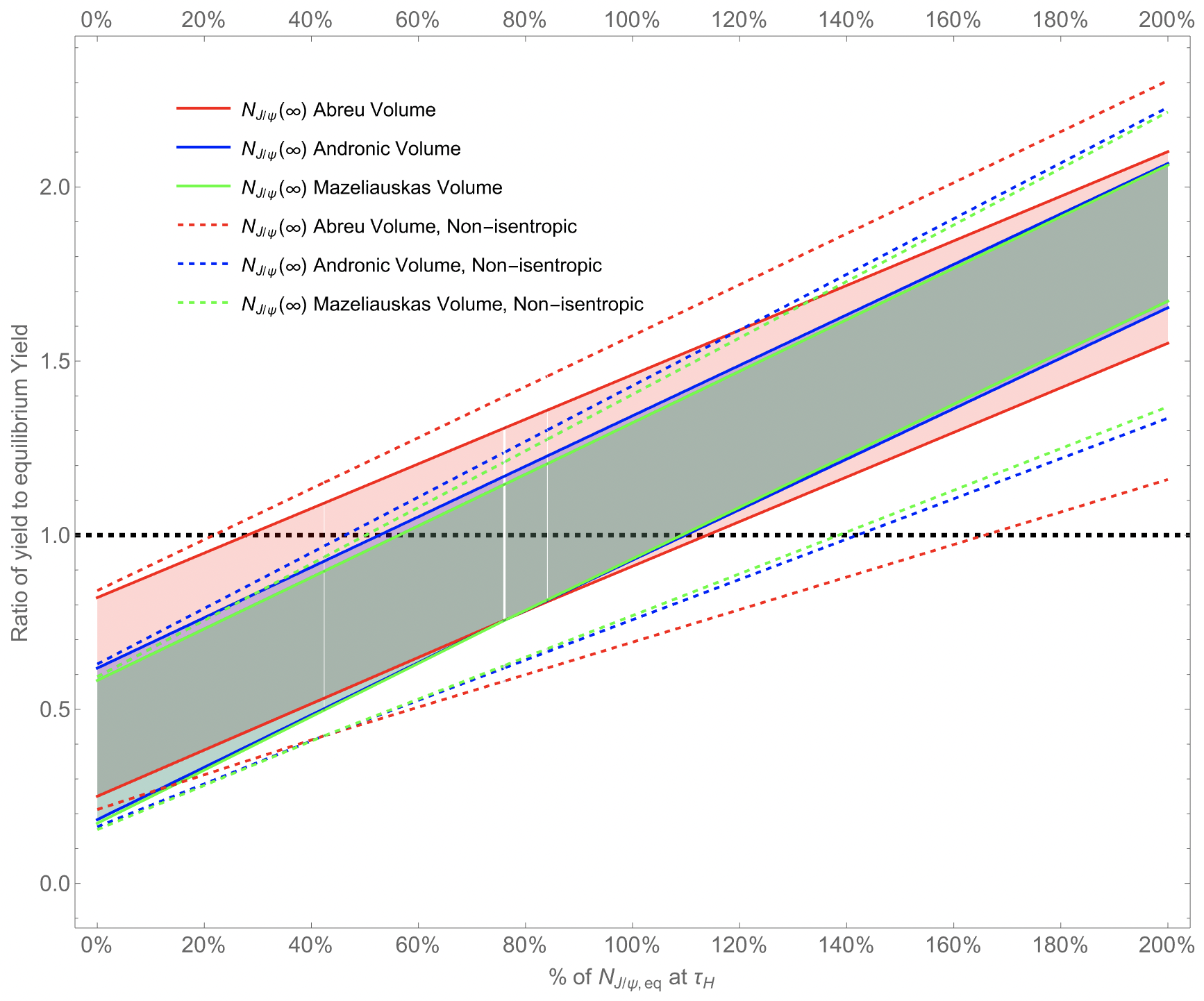}
    \caption{Ratio of total $J/\psi$ yield at $\tau=+\infty$ to equilibrium yield showing how the error bars grow if we relax the isentropic expansion assumption.}
    \label{fig:noniso}
\end{figure}

Can we relax the assumption of isentropic expansion? The only place the assumption of isentropic expansion factors into the final yield is in $A_{J/\psi}(\tau)$ which falls off rapidly until freeze-out time (which is calculated by assuming isentropic expansion) at which point it remains constant.\\
If we remain agnostic on how the temperature falls between $\sim 155 {\rm MeV}$ and $\sim 90 {\rm MeV}$, we can bound $A_{J/\psi}(\tau)$\\
\[155~{\rm MeV}\geq T(\tau\geq \tau_H) \geq 90~{\rm MeV}\]
This increases the error bars as shown in Fig.~\ref{fig:noniso} and makes the effect even more pronounced.

\subsection{Effective Model}

It is quite possible that the cross sections from the effective model are off by up to a factor of 2. We plot in Fig.~\ref{fig:factor2} a version of Fig.~\ref{fig:Percent} in which we incorporate this potential error:

\begin{figure}[H]
    \centering
    \includegraphics[width=0.8\linewidth]{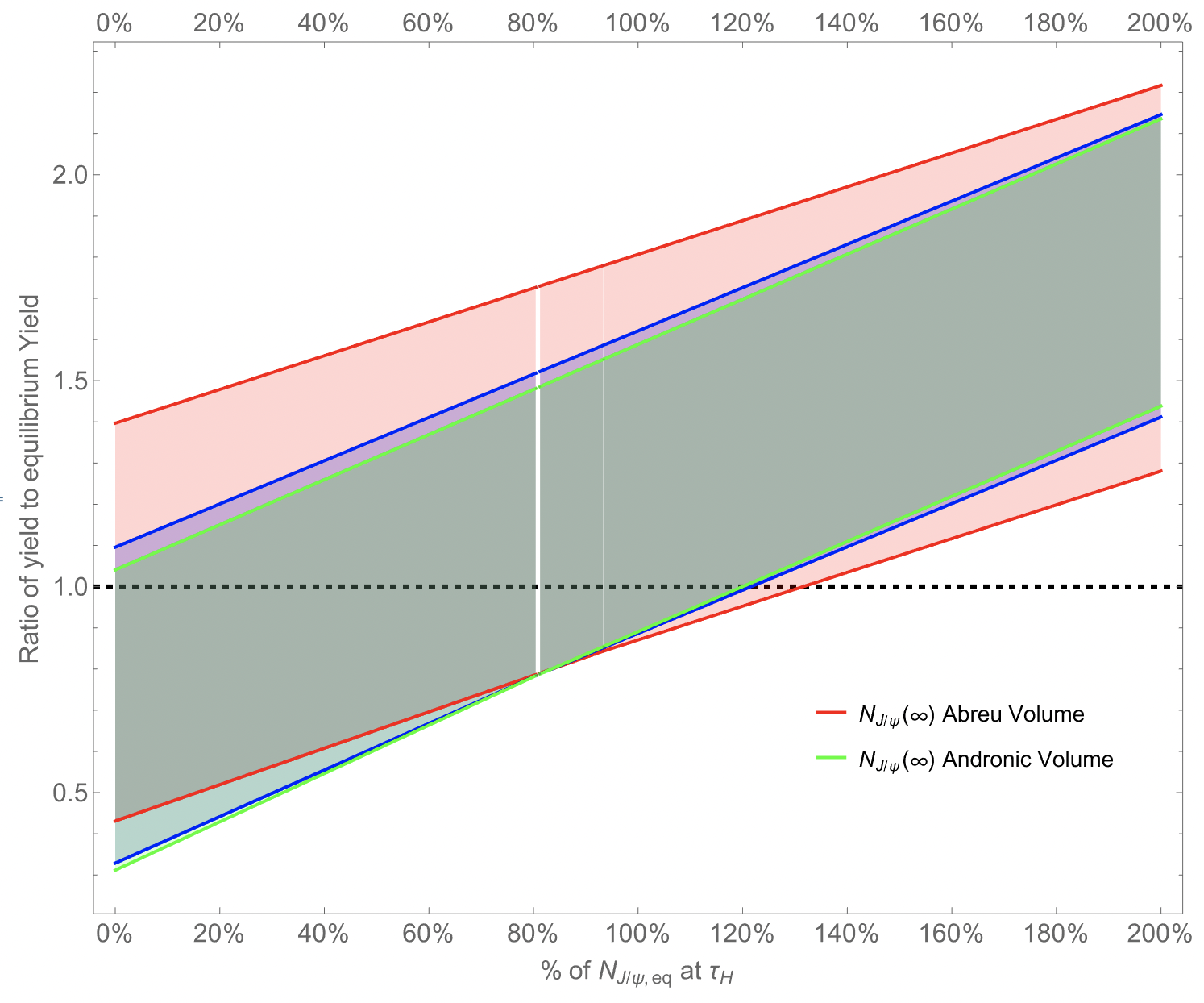}
    \caption{Ratio of total $J/\psi$ yield at $\tau=+\infty$ to equilibrium yield with $C_{J/\psi}'=2C_{J/\psi}$ and $A_{J/\psi}'(\tau)=2A_{J/\psi}(\tau)$. Systematic uncertainty assumed to be $10\%$ like (Figure \ref{fig:Percent})}
    \label{fig:factor2}
\end{figure}

If one assumes that the cross-sections are twice as large as the values quoted in \cite{Abreu:2017cof}, then it is consistent that as much as $100\%$ of total $J/\psi$ yield comes from hadronic regeneration.

\bibliographystyle{h-physrev5}
\bibliography{references}

\begin{thebibliography}{10}

\bibitem{ALICE:2012jsl}
ALICE, B.~Abelev {\em et~al.},
\newblock Phys. Rev. Lett. {\bf 109}, 072301 (2012), arXiv:1202.1383.

\bibitem{ALICE:2013osk}
ALICE, B.~B. Abelev {\em et~al.},
\newblock Phys. Lett. B {\bf 734}, 314 (2014), arXiv:1311.0214.

\bibitem{Braun-Munzinger:2000csl}
P.~Braun-Munzinger and J.~Stachel,
\newblock Phys. Lett. B {\bf 490}, 196 (2000), arXiv:nucl-th/0007059.

\bibitem{Thews:2000rj}
R.~L. Thews, M.~Schroedter, and J.~Rafelski,
\newblock Phys. Rev. C {\bf 63}, 054905 (2001), arXiv:hep-ph/0007323.

\bibitem{Andronic:2003zv}
A.~Andronic, P.~Braun-Munzinger, K.~Redlich, and J.~Stachel,
\newblock Phys. Lett. B {\bf 571}, 36 (2003), arXiv:nucl-th/0303036.

\bibitem{Andronic:2021erx}
A.~Andronic {\em et~al.},
\newblock JHEP {\bf 07}, 035 (2021), arXiv:2104.12754.

\bibitem{Andronic:2017pug}
A.~Andronic, P.~Braun-Munzinger, K.~Redlich, and J.~Stachel,
\newblock Nature {\bf 561}, 321 (2018), arXiv:1710.09425.

\bibitem{ALICE:2023gco}
ALICE, S.~Acharya {\em et~al.},
\newblock (2023), arXiv:2303.13361.

\bibitem{Datta:2003ww}
S.~Datta, F.~Karsch, P.~Petreczky, and I.~Wetzorke,
\newblock Phys. Rev. D {\bf 69}, 094507 (2004), arXiv:hep-lat/0312037.

\bibitem{KO1998237}
C.~Ko, X.~Wang, B.~Zhang, and X.~Zhang,
\newblock Phys. Lett. B {\bf 444}, 237 (1998).

\bibitem{Zhang:2002ug}
B.~Zhang, C.~M. Ko, B.-A. Li, Z.-W. Lin, and S.~Pal,
\newblock Phys. Rev. C {\bf 65}, 054909 (2002), arXiv:nucl-th/0201038.

\bibitem{Bratkovskaya:2003ux}
E.~L. Bratkovskaya, W.~Cassing, and H.~Stoecker,
\newblock Phys. Rev. C {\bf 67}, 054905 (2003), arXiv:nucl-th/0301083.

\bibitem{Linnyk:2008hp}
O.~Linnyk, E.~L. Bratkovskaya, and W.~Cassing,
\newblock Int. J. Mod. Phys. E {\bf 17}, 1367 (2008), arXiv:0808.1504.

\bibitem{2022136986}
ALICE, S.~Acharya {\em et~al.},
\newblock Physics Letters B {\bf 827}, 136986 (2022).

\bibitem{ALICE:2021rxa}
ALICE, S.~Acharya {\em et~al.},
\newblock JHEP {\bf 01}, 174 (2022), arXiv:2110.09420.

\bibitem{Abreu:2017cof}
L.~M. Abreu, K.~P. Khemchandani, A.~Mart\'\i{}nez~Torres, F.~S. Navarra, and
  M.~Nielsen,
\newblock Phys. Rev. C {\bf 97}, 044902 (2018), arXiv:1712.06019.

\bibitem{PhysRevC.51.2723}
K.~Martins, D.~Blaschke, and E.~Quack,
\newblock Phys. Rev. C {\bf 51}, 2723 (1995).

\bibitem{PhysRevC.62.045201}
C.-Y. Wong, E.~S. Swanson, and T.~Barnes,
\newblock Phys. Rev. C {\bf 62}, 045201 (2000).

\bibitem{Matinyan:1998cb}
S.~G. Matinyan and B.~Müller,
\newblock Phys. Rev. C {\bf 58}, 2994 (1998), arXiv:nucl-th/9806027.

\bibitem{Haglin:1999xs}
K.~L. Haglin,
\newblock Phys. Rev. C {\bf 61}, 031902 (2000), arXiv:nucl-th/9907034.

\bibitem{Lin:1999ad}
Y.-S. Oh, T.~Song, and S.~H. Lee,
\newblock Phys. Rev. C {\bf 63}, 034901 (2001), arXiv:nucl-th/0010064.

\bibitem{PhysRevC.62.034903}
Z.~Lin and C.~M. Ko,
\newblock Phys. Rev. C {\bf 62}, 034903 (2000).

\bibitem{Carvalho:2005et}
F.~Carvalho, F.~O. Duraes, F.~S. Navarra, and M.~Nielsen,
\newblock Phys. Rev. C {\bf 72}, 024902 (2005), arXiv:hep-ph/0508137.

\bibitem{ParticleDataGroup:2020ssz}
Particle Data Group, P.~A. Zyla {\em et~al.},
\newblock PTEP {\bf 2020}, 083C01 (2020).

\bibitem{Mazeliauskas:2019ifr}
A.~Mazeliauskas and V.~Vislavicius,
\newblock Phys. Rev. C {\bf 101}, 014910 (2020), arXiv:1907.11059.

\bibitem{Xu_2017}
J.~Xu and C.~M. Ko,
\newblock Phys. Lett. B {\bf 772}, 290 (2017).

\bibitem{ALICE:2013mez}
ALICE, B.~Abelev {\em et~al.},
\newblock Phys. Rev. C {\bf 88}, 044910 (2013), arXiv:1303.0737.

\end{thebibliography}

\end{document}